# Automated Generation of Dimensioned Rectangular Floorplans

Nitant Upasani, Krishnendra Shekhawat, Garv Sachdeva

Department of Mathematics, BITS Pilani, Pilani Campus, India-333031

**Abstract.** This paper proposes a methodology for the automated construction of rectangular floorplans (RFPs) while addressing dimensional constraints and adjacency relations. Here, adjacency relations are taken in the form of a dimensionless rectangular arrangement (RA) ensuring the existence of a RFP, while dimensional constraints are given in terms of minimum width and aspect ratio range for each room. A linear optimization model is then presented to obtain a feasible dimensioned RFP for user-defined constraints. A GUI is also developed for the automated generation of RFPs. The proposed model is able to generate feasible solutions for every possible RA in a reasonable amount of time.
From the architectural prospective, this work can be seen as a re-generation of well-known architectural plans with modified dimensions. At the end, the regeneration of existing legacy RFPs (corresponding to the user defined dimensions) has been demonstrated, taking their image as input.

## 1 Introduction

Floorplanning is among the essential steps of an architectural design program and has its applications in the compaction of electronic circuits as well. A *floorplan* is a partition of a finite-sized polygon $P$ into a finite set of dimensioned polygons $\{P_1, P_2 \ldots P_n\}$ called *rooms*. When the contained polygons $P_1, P_2 \ldots P_n$ are all internally disjoint rectangles and the envelope $P$ is convex, the floorplan is said to be a *rectangular floorplan* (RFP) with $n$ rooms. Ever since Levin [1] pointed out graph-theoretical correlations in the design of RFPs, several studies have made advancements showing a deeper relationship between the two.

It is stated that if $G(V, E)$ is an adjacency graph, then each element of the vertex set $V$ corresponds to a room and each edge $e_{ij} \epsilon E$, $E$ is the edge set, corresponds to an adjacency, i.e., a shared wall (or a section of the wall) between rooms $R_i$ and $R_j$ $(i, j \, \epsilon \, V)$ in the corresponding RFP, also known as the *rectangular dual* of graph $G$ (Figure 1(a)).

Designers are interested in constructing a RFP compatible with the adjacency relations provided by the graph; however, it is not always possible to do so (refer to Figure 1(b) which illustrates that the non-existence of a rectangular dual for a given adjacency graph $G$ implies the non-existence of a RFP for $G$; conversely, the existence of a rectangular dual does not guarantees the existence of a RFP because of the dimensional constraints). Hence in 1973, Steadman [2] proposed an algorithm to generate all topologically distinct rectangular duals for small values of $n$. Here, the problem of constructing a floorplan conforming to the adjacency graph was reduced to a problem of selecting an appropriate solution from the catalogue. Extending the idea further, in 1976 Mitchell et al. [3] provided a computer implementation of the algorithm, and proposed a model for dimensioning of the dual. However, apart from the restriction on the number of contained rectangles due to combinatorial considerations, the method proposed for dimensioning also had a drawback; it added extra constraints on the sizes of rooms. Roth et al. [4] were the first to give a step-by-step procedure for the construction of a dimensioned RFP from an adjacency graph, which did not depend upon search through a given catalogue. The adjacency graph was first split into two subgraphs using a colouring technique and extra connections were added to the graph for the same. This was followed by a series of steps before the two subgraphs were treated separately as flow networks to incorporate the dimensional constraints. But modeling of the network flow was such that it often led to the possibility of a loss of rectangularity in the solutions, i.e., either $P_1, P_2 \ldots P_n$ were not all rectangles, or the enclosure $P$ became non-convex. Later in 2000, Recuero et al. [5] presented an exhaustive set of heuristics to check whether extra connections are required to be added to a graph for the generation of its corresponding rectangular dual, but gave no proof for the same. After a thorough literature review we can claim that there is no algorithm in the literature which talks about the existence/non-existence of a rectangular dual for any arbitrary adjacency graph with certainty (there exist algorithms for constructing rectangular dual corresponding to planar triangulated graphs only [6, 7]). Hence, in this paper, we propose an alternative approach for representing adjacency relations; in the form of a *rectangular arrangement* ($RA$). A $RA$ is a dimensionless ordering of rectangular rooms such that the enclosure is a rectangle too. It can be seen as an RFP without dimensions. Hence, an approach of this kind guarantees the existence of a RFP conforming to the user-inserted adjacency requirements.

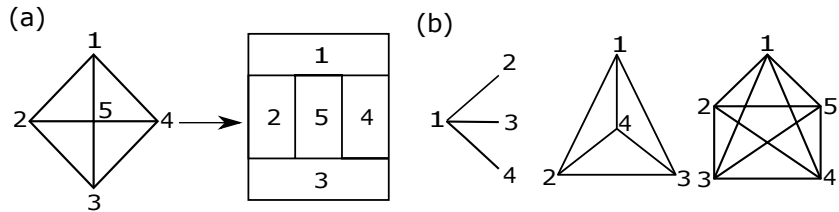

**Fig. 1.** (a) An adjacency graph and its rectangular dual. (b) Graphs for which rectangular dual and rectangular floor plan do not exist.

Some other notable studies are briefed in this paragraph. In 2010, Marson et al. [8] presented an algorithm for the construction of sliceable RFP such that the rooms are generated with an aspect ratio close to 1, given their areas. They also included circulations in the obtained RFP which are useful for easy access to each room. However, the objective of this study was not to preserve adjacency relations. In the same year, Merrel et al. [9] also gave a methodology for automated generation of building layouts for computer applications considering some thumb rules of architectural design for adjacencies and size of the rooms. In 2012, Eppstein et al. [10] introduced the concept of area-universal rectangular layouts and presented an algorithm for their construction from dual graphs, if such an arrangement exists. The peculiarity about these layouts was that any area could be assigned to each of the rooms and a feasible solution always existed. In 2013, Rodrigues et al. [11] developed an hybrid evolutionary approach for floorplan generation, enhanced with a local search technique called stochastic hill climbing (SHC) for complex dimensional and topological constraints. As a second part of [11], Rodrigues et al. [12] presented an implementation of their method, which although provided a feasible solution for highly detailed constraints, had a considerable amount of runtime and did not talk about retaining the rectangularity. In 2018, Shekhawat [13] introduced a concept of generic rectangular floorplans and proposed an algorithm for their enumeration, given the number of rooms. In the same year, Wang et al. [14] implemented a tool for customizing legacy floorplans by adding or removing a room using transformation rules. The floorplans, however, were restricted to RFPs whose dual graphs are planar triangulated graphs, and dimensional constraints were not considered either. Veloso et al. [15], in the same year, presented a sophisticated system for customized design of layouts, detailed as a Building Information Models (BIM), and established on the rule based shape-grammar paradigm for ready industrial use. More recently in 2019, Nisztuk et al. [16] developed an advanced application for the automated generation of floorplans satisfying all the adjacency and dimensional requirements but at the cost of rectangularity and additional spaces in between the rooms.

After a detailed literature review and to the best of our knowledge, there is no algorithm which addresses the dimensional constraints of a RFP while taking care of all the adjacency relations. In this paper, the problem is to develop a set of algorithms for constructing a RFP corresponding to given dimensional constraints while preserving the adjacency requirements, which are given in the form of a RA.

## 2 Preliminaries

In this section, we first define a few terminologies which are frequently used in this paper. These terms have been previously defined in the literature [2,7,17] and are reiterated here for a clear understanding.

**Definition 1.** Dual Graph of a RA

A *dual graph* is an undirected graph which can be constructed for any RA by replacing each room with a vertex and adding an edge to the vertices corresponding to the adjacent rooms (refer to Figure 2).

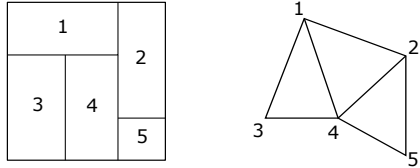

**Fig. 2.** Dual graph of a RA

**Definition 2.** ST graphs

An *st-graph* is a finite directed graph drawn without crossings in a plane, such that it has one source node $s$ (no incoming edges) and one sink node $t$ (no outgoing edges) on the exterior face and contains no directed cycle (refer to Figure 3(a)).

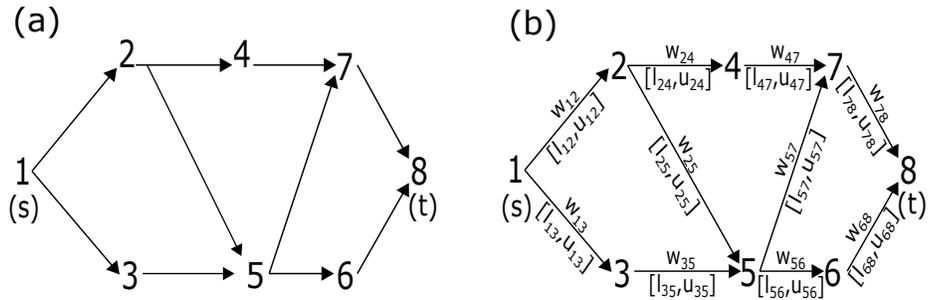

**Fig. 3.** (a) Example of an *st-graph* (b) Example of a network flow

**Definition 3.** Network flow

A *network flow* is defined as a weighted st-graph $G(V, E)$ such that each edge $e_{ij} \epsilon E$ receives a flow $w_{ij}$ greater than a lower bound $l_{ij}$ and less than a capacity $u_{ij}$ with the following restrictions: (i) the flow in each edge is non-negative (ii) the amount of flow into a node is equal to the amount of flow out of it, except for the source($s$) and sink($t$) nodes (refer to Figure 3(b)).

## 3 Methodology

In this section we discuss an algorithmic approach for the construction of a RFP for the given adjacency and dimensional constraints, where

(1) Adjacency constraints are given in the form of a RA.
(2) Dimensional constraints are given in the form of minimum width of each room along with the aspect ratio range for each room.

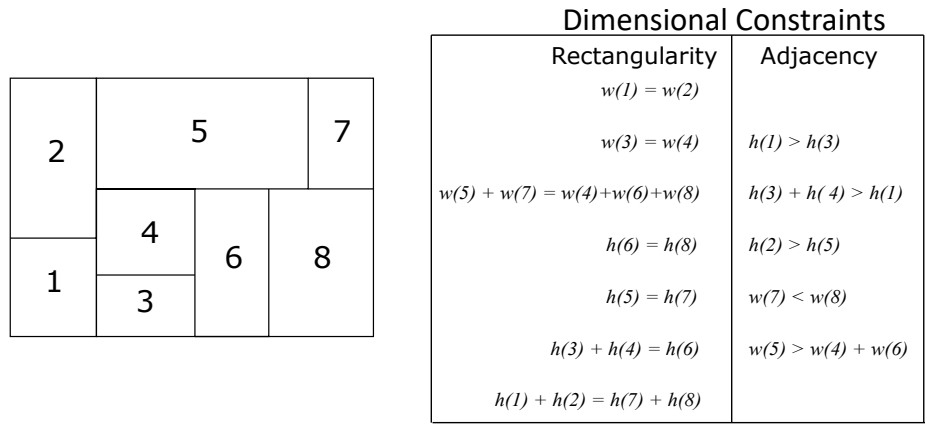

**Fig. 4.** RFP with 8 rooms and intrinsic dimensional constraints for preserving rectangularity and adjacency relations

The construction of a RFP from the given constraints is not straightforward. Additional dimensional constraints, other than the input, are required for preserving adjacency relations as well as rectangularity. For example, Figure 4 shows such additional constraints for a RA with 8 rooms. It can be clearly seen that mere listing of these additional constraints manually for all the rooms is a tedious task for even slightly higher values of $n$. In this paper, we not only generate all the constraints through our algorithm but also provide an optimization model to get a feasible RFP as the solution satisfying all these constraints along with the user-defined constraints. To proceed, a detailed methodology is presented below with the following major steps:

  i. Input a RA with dimensional constraints
 ii. Construction of *st-graphs*
iii. Flow network formulation
 iv. Optimization framework
  v. Solving for the RFP

For a better illustration, a RFP will be constructed step-by-step in this section, for the RA and dimensional constraints shown in figure 5. Keeping in mind the computer implementation, we will now discuss the steps of our methodology.

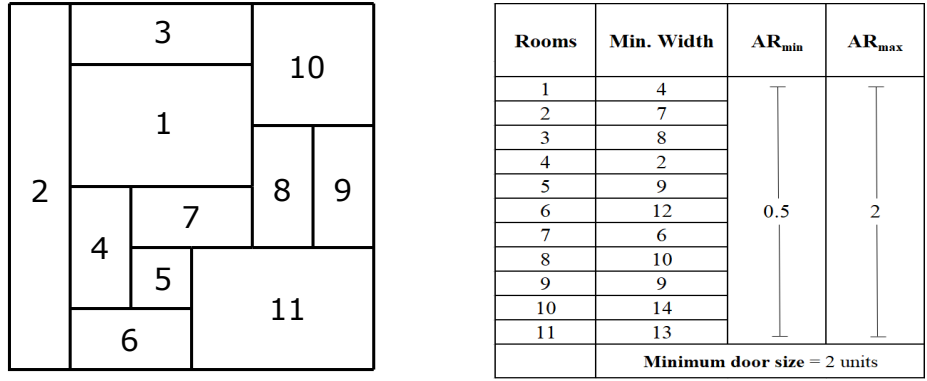

**Fig. 5.** Input RA with constraints

## 3.1 Input RA and dimensional constraints

For storing the RA in a computer memory, a data representation scheme is required. The data structure we use for storing RAs is a 2D matrix. It is not only an intuitive and simple way of storing data, but also reduces effort to extract relevant information like adjacency graphs, at the cost of space.

For its construction, the enveloping rectangle of the RA is divided into uniform rectangular grids such that every room is spanned by at least one grid element as shown in Figure 6. Each grid inside the enveloping rectangle is numbered

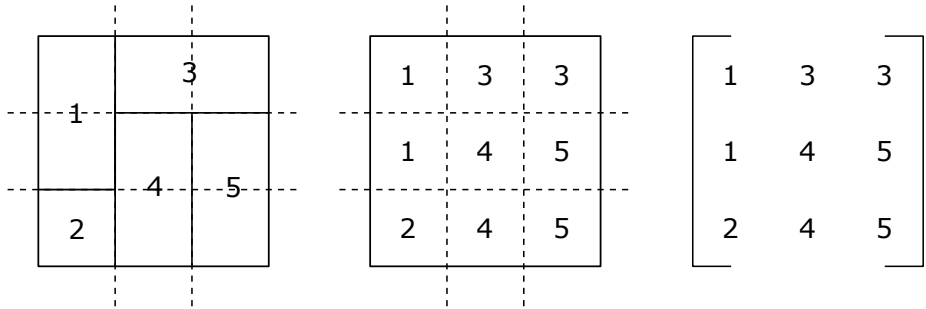

**Fig. 6.** Encoded Matrix

according to the index of the enclosing room, resulting in the *encoded matrix* (EM) of the RA.

## 3.2 Construction of ST-graphs

In this paper, we use st-graphs as a tool to extract adjacency relations from the input RA. These st-graphs, namely horizontal and vertical st-graphs (HST and VST) store adjacency relations in the X and Y direction respectively.
Corresponding to the input rectangular arrangement, HST and VST will have $n+2$ vertices each, where each room is represented by a vertex, and the two additional vertices, correspond to the source and sink vertices. The steps for the construction of both *st-graphs* are given below and shown in Figure 7.

i. The first step is to add a rectangle to each side of the RA such that the overall composition is rectangular, as shown in Figure 7(a). The EM is updated accordingly.
ii. Next, a dual graph of the given RA is constructed, as shown in Figure 7(b).
iii. The dual graph is split into two graphs based on the horizontal and vertical adjacencies. They are referred to as horizontal adjacency graph (HAG) and vertical adjacency graphs (VAG) respectively.
iv. Source and sink nodes are assigned for converting the undirected HAG and VAG to directed st-graphs. North(N)/South(S) are assigned as source/sink for VAG and West(W)/East(E) as source/sink for HAG. According to the definition of st-graphs, all the edges connected to the source nodes are outgoing whereas those connected to the sink nodes are incoming. The rest of the edges are directed from top to bottom for VAG and left to right for HAG.

The resulting directed HAG and VAG are the two required st-graphs HST and VST respectively (Figure 7d). From a programming perspective, we can obtain the HST and VST directly from the EM (Figure 7(c)) using the following steps.

**Step 1**: For HST, each row of the EM is traversed (left to right), and for every distinct pair of consecutive entries, a directed edge is added between their corresponding vertices. The direction of the edges is from the former entry to the latter. However, the vertices North (N) and South (S) are deleted from the HST.
**Step 2**: For VST, each column of the EM is traversed (top to bottom), and for every distinct pair of consecutive entries, a directed edge is added between their corresponding vertices. The direction of the edges is from the former entry to the latter. However, the vertices West (W) and East (E) are deleted from the HST.

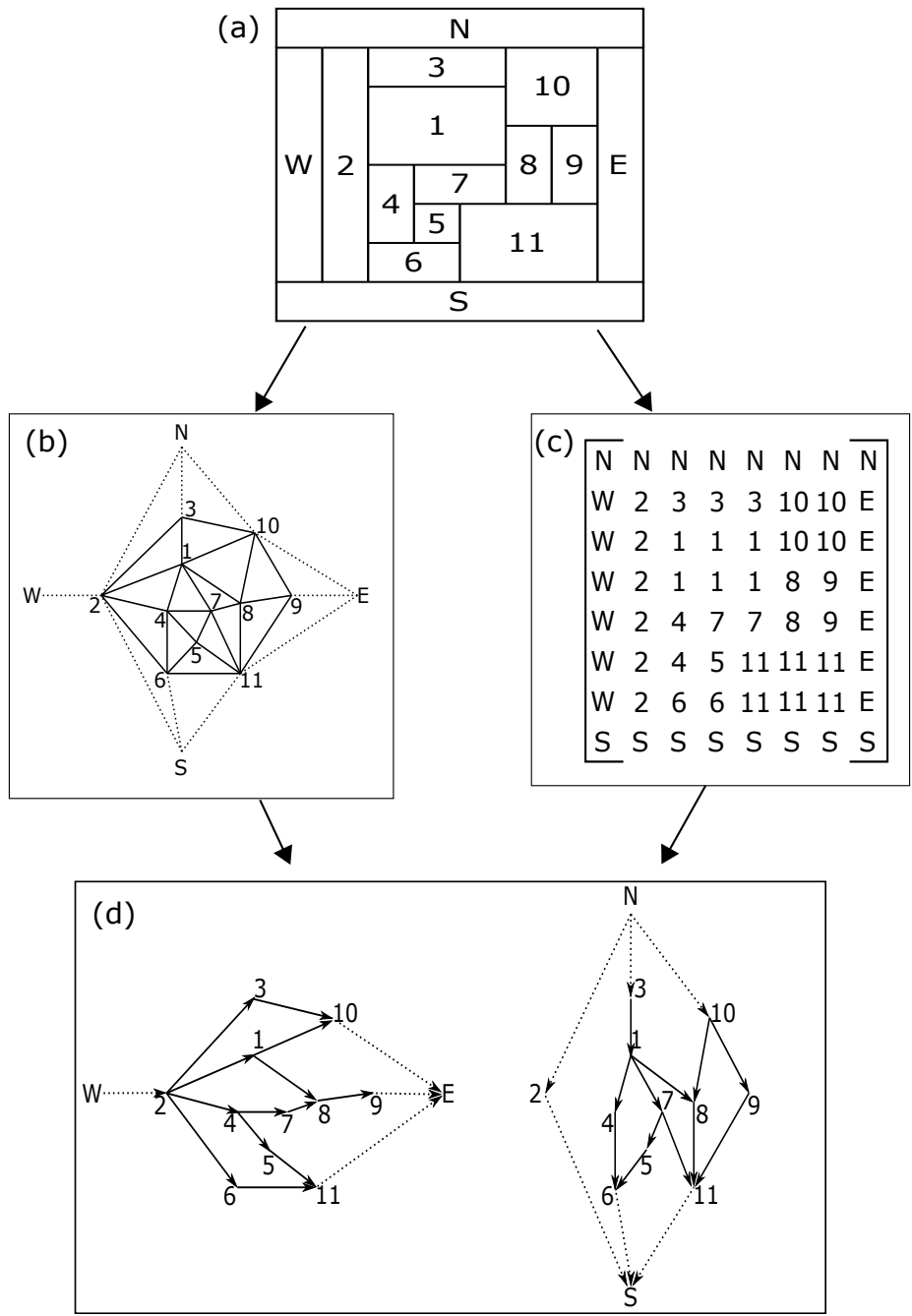

**Fig. 7.** Worked out example: Construction of *st-graphs*

### 3.3 Flow network formulation

The HST and VST are intuitive ways for representing the respective adjacencies. This intuitiveness can further be extended by modelling the st-graphs as flow networks, where each edge carries a non-zero, non-negative flow. Corresponding to the unweighted HST and VST, a weighted HST (HNF) and VST (VNF) are designed as network flows (refer to Figure 8). For a proper correlation between the st-graphs and flow networks, the following assumptions have been made:

i. The amount of flow through each edge represents the magnitude of wall section common to two rooms in the respective RFP.
ii. In HNF/VNF, the amount of inward or outward flow in a node is a measure of the height/width of the room corresponding to the node.

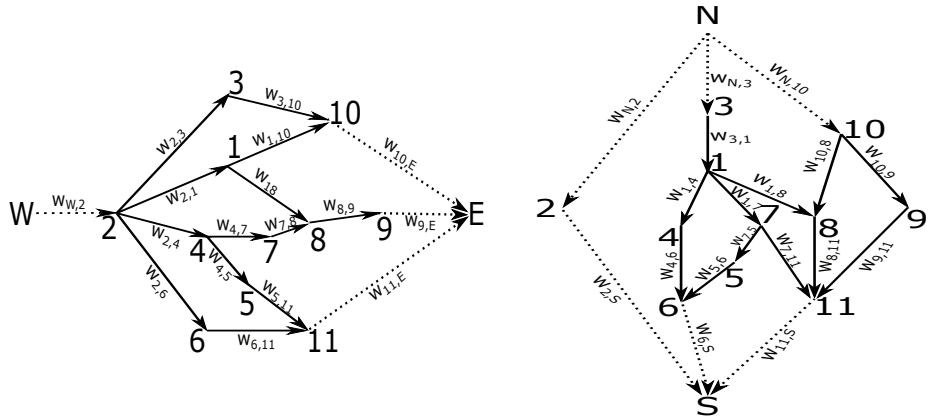

**Fig. 8.** HNF and VNF for worked out example

Before moving on to the optimization framework, a few redundant edges in the directed st-graphs may also be removed. Since the inflow is equal to the outflow through any node, edges entering the nodes south and east, can be removed safely.

### 3.4 Optimization Framework

Linear optimization techniques have been used on the two obtained flow networks to incorporate the dimensional constraints and to obtain a feasible solution of minimum area satisfying the given adjacencies. The framework essentially remains the same for VNF as well as HNF and is explained below.

*Design Variables*

Let *G(V,E)* denotes the network flow that is being considered, and $m$ be the cardinality of the directed edge set *E*.

Corresponding to each directed edge, we have a design variable. Hence the total number of design variables are $m$.

*Objective Function*

The objective function chosen is such that it minimizes the total dimension of the rectangular floorplan corresponding to the respective flow network. That is, VNF will minimize the overall width whereas HNF will minimize the total height of the floorplan. The objective function $f$ is defined as

$$f = \sum w(e_{sj}) \tag{1}$$

where $w(\cdot)$ denotes the weight function, $e_{sj}$ denotes the directed edge from node $s$ to node $j$, and $s \in (N, W)$ is the source node.

*Design constraints*

Equality constraints: The equation for equality constraints are obtained from the condition stated in assumption (2) in Definition 3.

$$\sum w(e_{ji}) = \sum w(e_{ik}) \qquad \forall i \in V(G) \tag{2}$$

where $\sum w(e_{ji})$ denotes the total inflow and $\sum w(e_{ik})$ denotes the total outflow from node $i$.

Inequality constraints: The minimum and maximum dimensional constraints of each room will be used as the inequality constraints.

$$\min(d) \leq \sum w(e_{ik}) \leq \max(d) \qquad \forall i \in V(G) \tag{3}$$

where $d$ is the dimension corresponding to flow network in consideration, i.e., width for VNF and height for HNF.

*Bounds for the design variables*

Each design variable corresponds to the magnitude of wall section common to the two rooms and hence a lower bound for the size of the door is selected for the framework. The algorithm is also capable of handling multiple door sizes for distinct room connections.

### 3.5 Solving for the RFP

A series of steps have been proposed in this section which involves solving VNF/HNF using linear optimization techniques multiple times. It should be noted that for a given flow network, the design variables, objective function and equality constraints always remain the same, and only the inequality constraints, i.e., minimum dimensional constraints govern the optimization process.

***Step 1:*** *Input dimensional constraints*

Dimensional constraints are taken as an input from the user in the form of minimum width ($w_{i,min}$) and the permitted aspect ratio range ($AR_{i,min}$, $AR_{i,max}$) for each room $i$. Figure 5 shows the dimensional constraints of the solved example in consideration.

***Step 2:*** *Solving for width*

The width constraints are taken as inequality constraints from the input for optimization on VNF, and solved for a minimum width of the rectangular floor-plan using simplex method, however, any other linear optimization technique may be used. Hence an optimal solution of minimum width (of the complete RFP), giving values of each design variable, is obtained. The width of each room $w_i$ is calculated by summing the total inflow ($\sum w(e_{ji})$) towards a room.

***Step 3:*** *Solving for height*

Width of rooms ($w_i$) is multiplied by lower limit of the aspect ratio ($AR_{i,min}$) to get the minimum height ($h_{i,min}$) of each room.

$$h_{i,min} = w_i \times AR_{i,min} \qquad (4)$$

Considering the minimum height of each room as inequality constraints, HNF is solved using a linear optimization technique to get height of each room (similar to step 2). Since the total height of the RFP (objective function) is to be minimized, not inserting the maximum height constraints of each room would not render the optimization problem unbounded.

***Step 4:*** *Check for aspect ratio bounds and update minimum width*

After step 3, the initial dimensions ($w_i$, $h_{i,min}$) of each room are obtained. The aspect ratio of some of the rooms however may exceed the permissible value ($AR_{i,max}$) considering that the height constraints were not bounded. Since the initial $h_{i,min}$ of these rooms, calculated using $w_i$, give an aspect ratio beyond

$AR_{i,max}$, $w_{i,min}$ is updated using the following equation.

$$w_{i,min(up)} = h_{i,min}/AR_{i,max} \qquad (5)$$

Once the minimum width is updated, steps 1-4 are repeated till the aspect ratios of all the rooms lie in between the specified range. The final updated heights and widths are the final feasible dimensions of the rooms.

***Step 5:*** *Constructing the floorplan*

After getting the final dimensions of all the rooms, the RFP is constructed using the EM. The columns are traversed from left to right, and the rooms are constructed in order of occurrence of their respective entries.

For a better understanding of how the proposed optimization model converges to a feasible solution, Figure 9 illustrates intermediate RFPs obtained from each iteration of the steps in this subsection.

## 4 Results

A prototype has been developed as an implementation of the algorithm presented in this paper, using MATLAB R2018a. It prompts a GUI for the user to draw any desirable RA, and takes dimensional constraints for the rooms as input (refer to Figure 10). Since linear optimization has been used to obtain a feasible solution, the algorithm takes few seconds to produce a feasible RFP. Figure 10(c) shows the final RFP with dimensions written in parenthesis under each room index. It can be seen that the adjacency relations are retained in the dimensioned solution, providing wall section of at least a minimum width (decided by the user) to accommodate doors between the adjacent rooms. Moreover, the position (geometry) of the rooms as in the RA is also retained in the final RFP.

From an architectural perspective, the RA provided by the user may not have an architectural meaning. Hence, this work can be seen as a re-generation of architecturally well-known rectangular floorplans (RFPs) for user-defined dimensional constraints. The adjacency relations, as well as aspect ratios inferred from existing RFPs have architectural significance and must be preserved in the regenerated solution along with the user-constraints.

Here, designers can give images of existing RFPs as input to regenerate a solution conforming to user-defined dimensional constraints. First, the adjacency relations are extracted from these images with the help of image processing tools, to construct a dual graph of the existing RFP. Then the aspect ratios are extracted from the input image for defining an approximate range for each room (adhering to these aspect ratios has a significant meaning here, as certain rectangular spaces, such as circulations, have a typically distinct aspect ratio which must not be altered from an architectural and aesthetic viewpoint).

Figure 11 shows the image of a well-known floorplan, i.e. Palladio floorplan

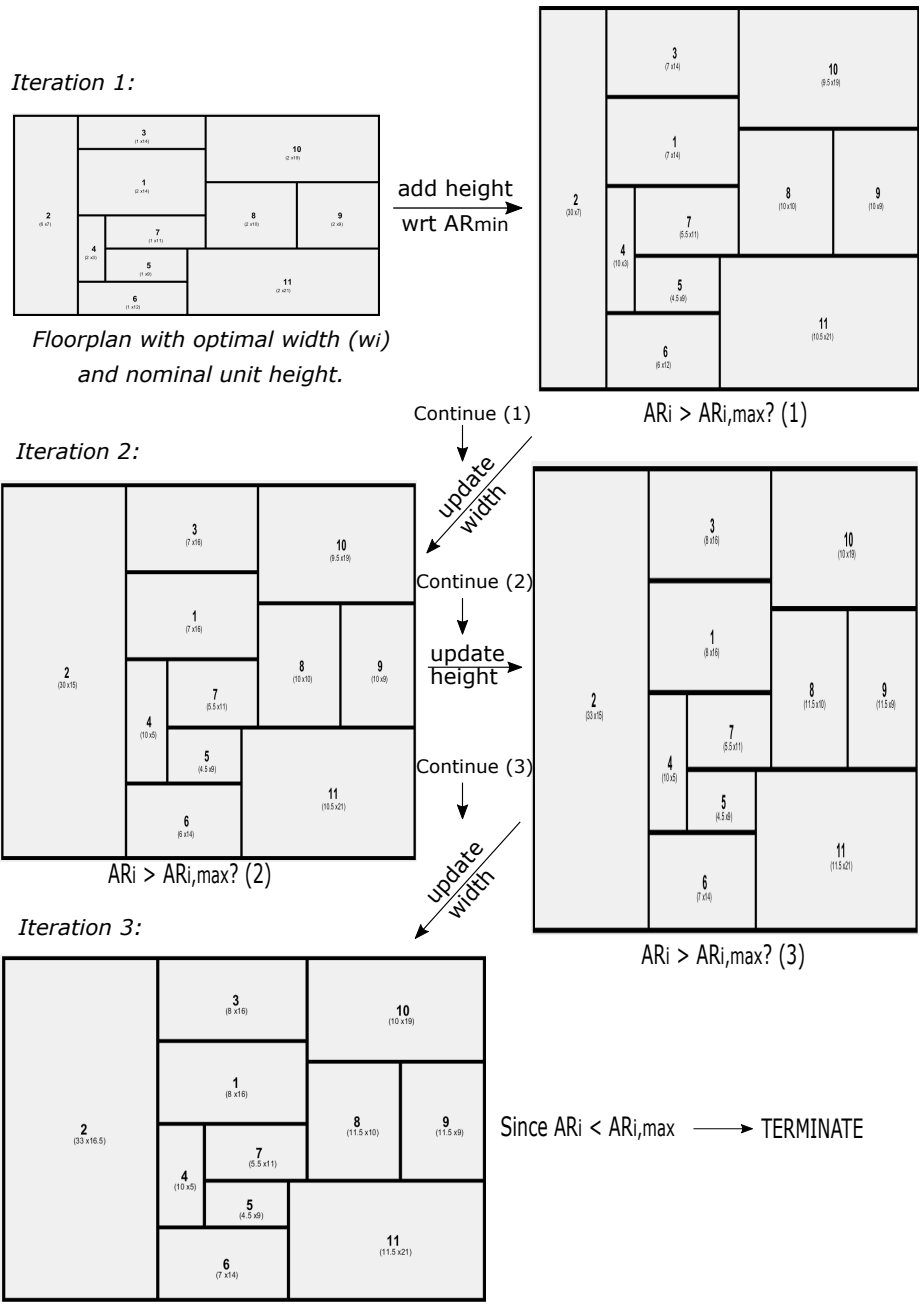

**Fig. 9.** Step-by-step convergence of optimization

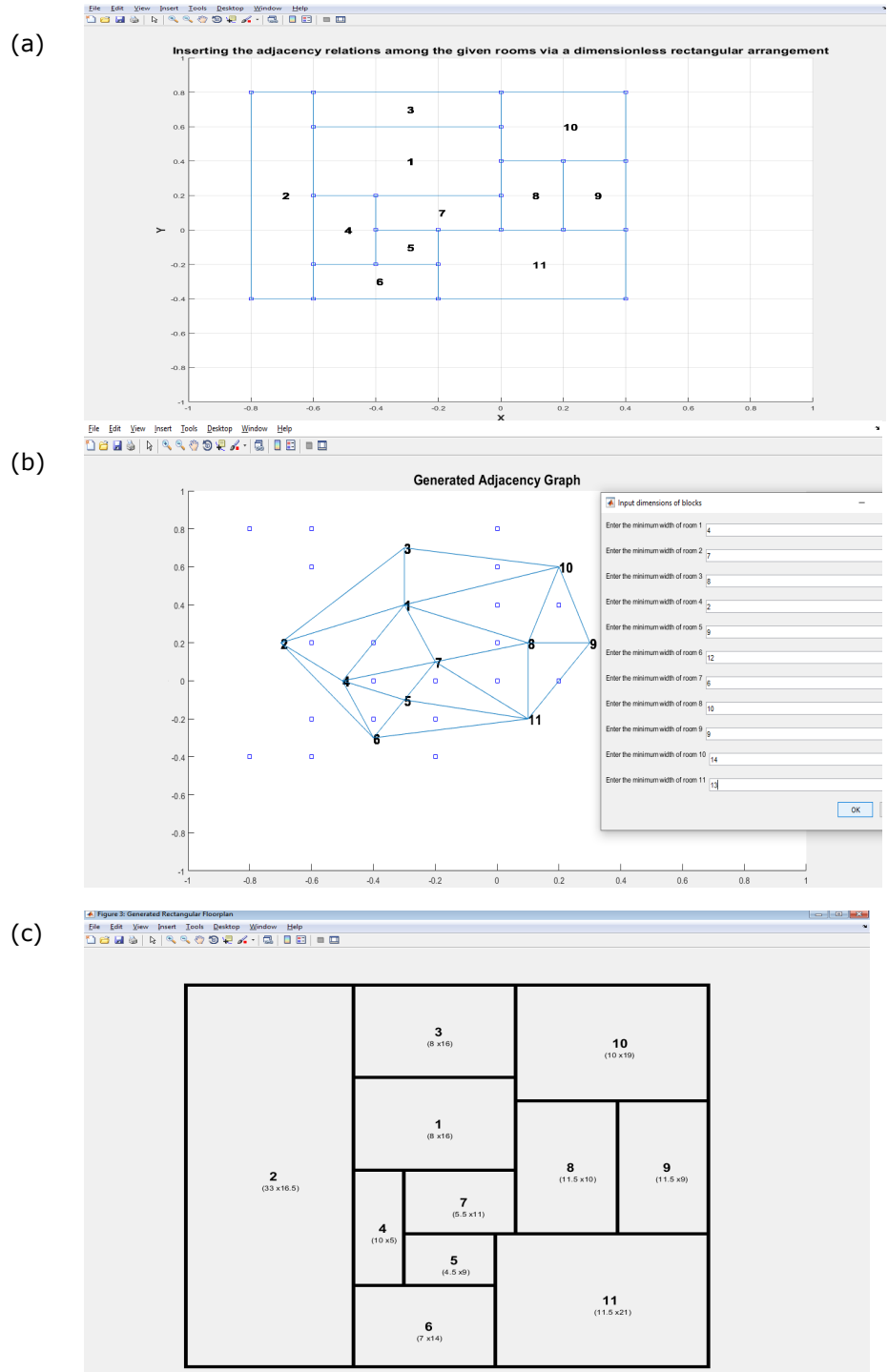

**Fig. 10.** Interface of the prototype (a) Input RA (b) Input dimensional constraints (c) Final generated RFP

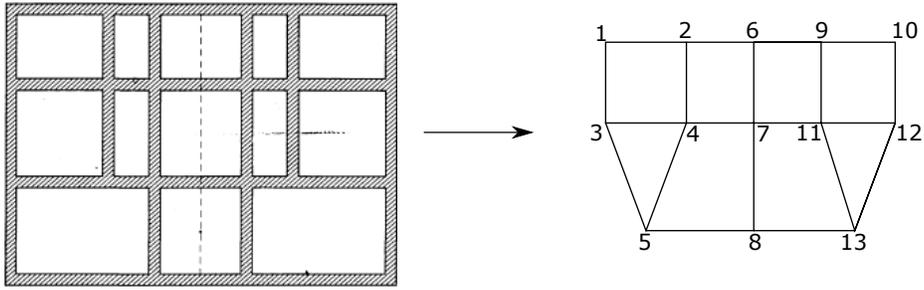

**Fig. 11.** (a) Input image of renowned Palladio floorplan (image from Stingy et al. [18]) and its adjacency graph as extracted from the image

[18] and its extracted adjacency graph. The adjacency graph and aspect ratios of rooms derived from the image are used to regenerate the RFP using the network-flow optimization algorithm proposed in Section 3. The regenerated RFP (Figure 12) is in compliance with the dimensional constraints shown alongside, as an example.

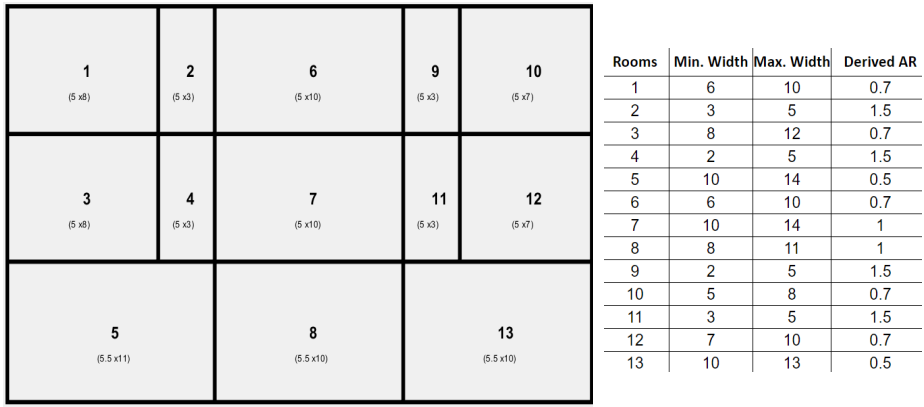

| Rooms | Min. Width | Max. Width | Derived AR |
|---|---|---|---|
| 1 | 6 | 10 | 0.7 |
| 2 | 3 | 5 | 1.5 |
| 3 | 8 | 12 | 0.7 |
| 4 | 2 | 5 | 1.5 |
| 5 | 10 | 14 | 0.5 |
| 6 | 6 | 10 | 0.7 |
| 7 | 10 | 14 | 1 |
| 8 | 8 | 11 | 1 |
| 9 | 2 | 5 | 1.5 |
| 10 | 5 | 8 | 0.7 |
| 11 | 3 | 5 | 1.5 |
| 12 | 7 | 10 | 0.7 |
| 13 | 10 | 13 | 0.5 |

**Fig. 12.** Regenerated Palladio RFP with new dimensions

## 5 Conclusions and Future Work

This paper presents an algorithmic approach for the generation of feasible RFPs, preserving adjacency relations extracted from a given RA and satisfying dimensional constraints provided by the user. In addition, the RFPs generated are not only unrestricted to sliceable RFPs [8] or to the RFPs whose dual graph is a planar triangulated graph [14], but cover all those RFPs which contain four rooms meeting at a common vertex, which is a rarely discussed issue in the literature. To the best of our knowledge, existing literature does not provide an algorithm for the construction of a RFP while preserving adjacencies and simultaneously incorporating dimensional constraints, and we are the first to provide a detailed methodology for the same.

The concept of regenerating existing RFPs, presented towards the end of this paper, is eminently essential for architects, as they generally have a catalogue of floorplan designs to choose from but need assistance in dimensioning, abiding the user-requirements and then choosing the most appropriate design.

The future work has two major facets which can be seen as follows:

1. Mathematically, the final RFP derived is programmed to have an area as small as possible, but no comments have been made about the optimality of the obtained solution. Studies in the literature have employed non-linear optimization techniques for reducing the total area of the floorplan, however, no one has come close to provide an algorithm for finding the optimal solution. In future, the derivation of such an algorithm by extending the network flow optimization model looks promising.
2. Architecturally, the restriction to rectangularity can be eliminated by considering orthogonal floorplans, which will further broaden the scope of the study to a majority of the floorplans. The challenge in its implementation lies not in transforming the orthogonal floorplans to RFPs, but in incorporating the constraints of these orthogonal rooms in the network flow model.

Moreover, the prototype developed as a proof-of-concept in this paper will be integrated with softwares like Rhinoceros and Grasshopper, in the future, pertaining to its importance in the architectural industry.

## Acknowledgement

The research described in this paper evolved as part of the research project Mathematics-aided Architectural Design Layouts (File Number: ECR/2017/000356) funded by the Science and Engineering Research Board, India.